\documentclass{article}
\usepackage{spconf,amsmath,graphicx,hyperref}

\usepackage{amsmath,graphicx,flushend}
\usepackage{indentfirst}
\usepackage{dblfloatfix}
\usepackage{tikz,hyperref,makecell}
\usetikzlibrary{positioning}
\usepackage{diagbox}
\usepackage{soul}
\usepackage{caption}
\usepackage{multirow}
\usepackage{listings}
\usepackage{xcolor}
\usepackage{hyperref}
\usepackage{mathtools}
\usepackage{booktabs}
\definecolor{codegreen}{rgb}{0,0.6,0}
\definecolor{codegray}{rgb}{0.5,0.5,0.5}
\definecolor{codepurple}{rgb}{0.58,0,0.82}
\definecolor{backcolour}{rgb}{0.95,0.95,0.92}
\lstdefinestyle{mystyle}{
    backgroundcolor=\color{backcolour},   
    commentstyle=\color{codepurple},
    morecomment=[l][\color{codegreen}]{//},
    keywordstyle=\color{magenta},
    morekeywords={function}
    numberstyle=\tiny\color{codegray},
    stringstyle=\color{codegreen},
    basicstyle=\ttfamily\footnotesize,
    breakatwhitespace=false,         
    breaklines=false,                 
    captionpos=b,                    
    keepspaces=false,                 
    numbers=left,                    
    numbersep=2pt,  
    xleftmargin=1em,
    showspaces=false,                
    showstringspaces=false,
    showtabs=false,                  
    tabsize=1
}

\def\BibTeX{{\rm B\kern-.05em{\sc i\kern-.025em b}\kern-.08em
    T\kern-.1667em\lower.7ex\hbox{E}\kern-.125emX}}
\title{Room Impulse Response Synthesis via Differentiable Feedback Delay Networks for Efficient Spatial Audio Rendering}

\name{Armin Gerami and Ramani Duraiswami\thanks{Supported by ONR Award N00014-23-1-2086.}}
\address{Perceptual Interfaces \& Reality Lab\\ Department of Computer Science \& UMIACS\\ University of Maryland,  College Park}
\begin{document}
\topmargin=0mm
\ninept
\maketitle
\begin{abstract}
We introduce a computationally efficient and tunable feedback delay network (FDN) architecture for real-time room impulse response (RIR) rendering that addresses the computational and latency challenges inherent in traditional convolution and Fourier transform based methods. Our approach directly optimizes FDN parameters to match target RIR acoustic and psychoacoustic metrics such as clarity and definition through novel differentiable programming-based optimization. Our method enables dynamic, real-time  adjustments of room impulse responses that accommodates listener and source movement. When combined with previous work on representation of head-related impulse responses via infinite impulse responses, an efficient rendering of auditory objects is possible when the HRIR and RIR are known. Our method produces renderings with quality similar to convolution with long binaural room impulse response (BRIR) filters, but at a fraction of the computational cost.
\end{abstract}


\vspace*{0pt}
\section{Introduction}
\label{sec:intro}
Binaural Room Impulse Response (BRIR) is a central component of modern spatial audio rendering, which aims to recreate 3D soundscapes over headphones. With the increasing popularity of personalized augmented reality (AR) and virtual reality (VR) devices, creating a realistic and personalized BRIR on the fly, for a listener moving in a real or virtual world relative to the sound objects in the scene, is crucial for creating immersive auditory experiences. A BRIR is created by convolution of two filters and carries two sets of acoustic cues: those related to the room characteristics through the Room Impulse Response (RIR) and those related to scattering of the listener's anatomy through the Head-Related Impulse Responses (HRIRs). The tail of the RIR imparts  the audible acoustic characteristics of a physical space, arising from sound interacting with surfaces through reflections, diffraction, and absorption. The direct sound and the early reflections in the RIR, coming within the first 50 to 100 ms, provide location information and aid intelligibility. 

\par In practice, spatial audio systems apply these impulse responses to ``dry" audio signals, typically using either direct convolution-based methods for shorter filters, or Fast Fourier Transform (FFT) based approaches, to enhance realism in gaming, VR, AR, and immersive headphone listening of media. However, these methods present challenges in balancing computational efficiency, perceptual accuracy, and low latency, especially on edge wearable devices with size, weight, and power constraints. The situation is more challenging when the impulse responses must be continuously adapted to account for moving listeners/sources or changing environments \cite{zotkin2004rendering}. In our previous work we addressed the efficient application of HRIRs~\cite{gerami2025efficient}, achieving a threefold improvement in computational time, and fivefold improvements in both memory and latency reduction. This paper focuses on developing a computationally efficient and adaptable model for the RIR component.

\par Recent work in spatial audio has focused on leveraging data-driven methods to address these challenges. For instance, machine learning frameworks, such as neural networks trained on large RIR datasets, have attempted to synthesize spatially accurate acoustic fields ~\cite{luo2022learning, zhong2021acousnet, sanaguano2022deep, chen2024real, mckenzie2022perceptually, shen2020data, pezzoli2023implicit, karakonstantis2023room, sharma2025machines}. Concurrently, parametric approaches that decompose RIR and HRIRs into perceptually salient components have been proposed, where the parameters are learned through differentiable programming~\cite{gerami2025efficient, masuyama2024niirf, zhi2023differentiable, Jot1993-pm}.

\par Despite these advances, existing systems struggle to reconcile the computational demands of high-fidelity convolution, the delay introduced by the Fourier transform interferes with the latency constraints of real-time perception; or of real-time adaptability. We propose a computationally lightweight, feedback delay network (FDN) for RIR rendering that addresses these. Our approach aims to capture the precise desired acoustic and psychoacoustic metrics such as clarity and definition, while maintaining computational efficiency for real-time applications. We present a differentiable programming-based optimization that ensures a solution for the FDN parameters that produces an RIR rendering with the desired characteristics. We also suggest a rendering framework for BRIRs, that incorporates our approach for RIRs, and the efficient differentiable IIR matching approach suggested in \cite{gerami2025efficient} for HRIRs. In contrast to previous work, we do not rely on real RIR measurements or simulations, which are difficult to acquire or compute. Instead, we directly use perceptual acoustic and psychoacoustic metrics during optimization. Furthermore, the FDN parameters can be updated real-time, accommodating listener/source movement. By evaluating both objective metrics and subjective listener assessments, we demonstrate our approach.

\vspace*{0pt}
\section{Background}
\label{sec:back}
\begin{figure}[h] 
  \centering
  \includegraphics[width=\columnwidth,trim=0 0pt 0pt 0, clip]{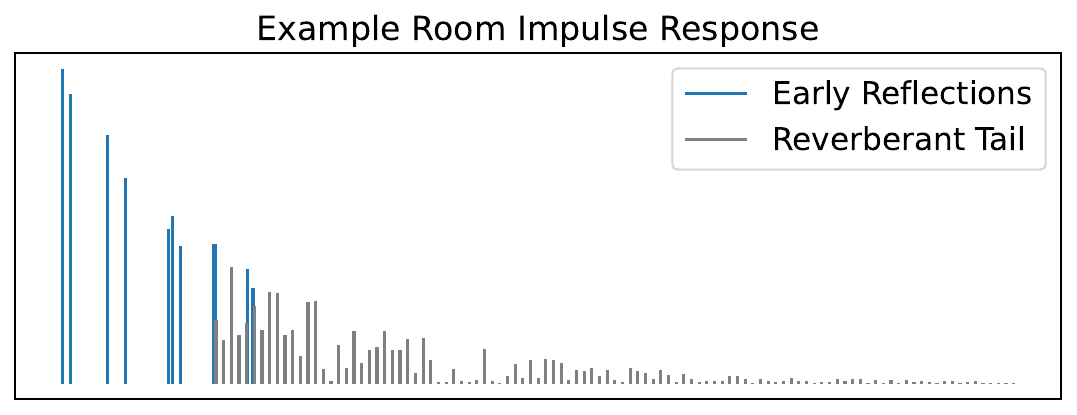} 
  \caption{Example room impulse response partitioned into the early reflections and the reverberant tail segments.}
  \label{fig:rir} 
\end{figure}

\vspace*{0pt}
\textbf{Room Impulse Responses}\  (see Fig.~\ref{fig:rir}) can be segmented along the time axis into  early (1st- and 2nd-order) reflections and the reverberant tail. For a shoebox room, the early reflections comprise 43 coefficients (1 for direct path, 6 for 1st-order, and 36 for 2nd-order reflections). For small rooms and large conference halls, this spans the first 50 ms to 250 ms. Given a sampling rate of $\sim$48 kHz, the early reflections in a RIR are  very sparse. The denser reverberant tail encompasses all higher-order reflections and can last from several from hundreds of ms to a few seconds.
\par Applying an RIR of length $N$ to a signal  can be done either via $O(N^2)$ cost time-domain convolution, or $O(N \log N)$ frequency domain convolution.  The time-domain approach is costly due to the RIR's length ($\sim10^4$ samples), whereas the more efficient frequency-domain approach,  introduces latency due to the trade-off between Fourier transform accuracy and window size. Methods to mitigate these by using various partitioned convolution approaches have been proposed \cite{Torger2002-eo, zotkin2004rendering}.

\par \textbf{Acoustic Metrics:} Human perception of spatial audio in a room is shaped by both the early reflections and the reverberant tail. While we are sensitive to the specific values of the early reflections,  our perception of the reverberant tail is well characterized by the average  psychoacoustic metrics~\cite{kuttruff2016room, neidhardt2022perceptual} below.
\begin{itemize}
    \item Clarity ($C$): $C = \log\left({\int_0^{50\text{ms}}f^2(t)\,dt}/{\int_0^{\infty}f^2(t)\,dt}\right)$\vspace{2pt}
    \item Definition ($D$): $D = {\int_0^{80\text{ms}}f^2(t)\,dt}/{\int_0^{\infty}f^2(t)\,dt}$\vspace{2pt}
    \item Center Time ($CT$): $CT = {\int_0^{\infty}tf^2(t)\,dt}/{\int_0^{\infty}f^2(t)\,dt}$\vspace{3pt}
    \item $T_{30}$: The time it takes for a sound to decay by 30 dB.
\end{itemize}
\par We develop a novel approach to specify a FDN that produces the same results as a convolutional RIR as far as the acoustic and psychoacoustics characteristics are concerned, while significantly reducing computational costs and without introducing latency.
\begin{figure}[htb] 
  \centering \includegraphics[width=0.65\columnwidth,trim=0 160pt 380pt 0, clip]{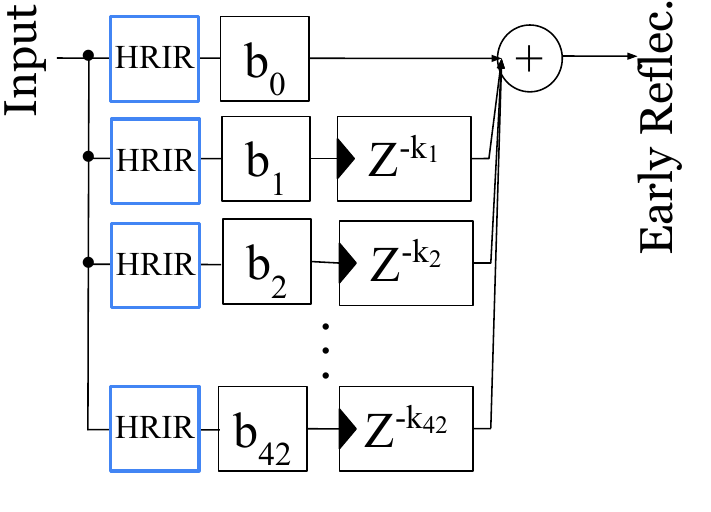}   \includegraphics[width=0.85\columnwidth,trim=0 40pt 300pt -15pt, clip]{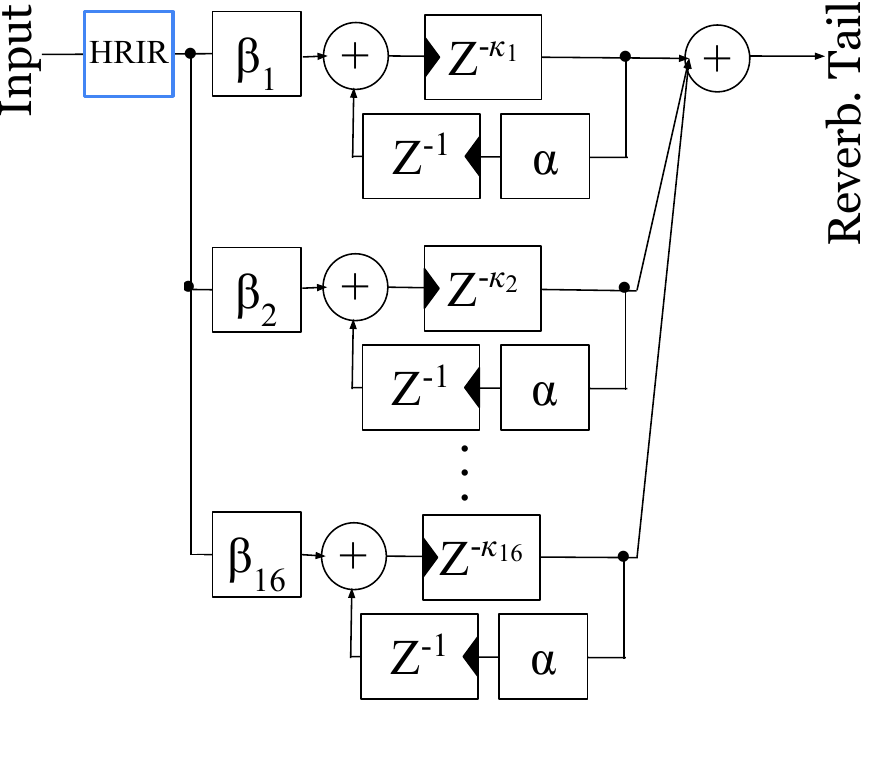}  \includegraphics[width=0.65\columnwidth,trim=0 260pt 340pt -15pt, clip]{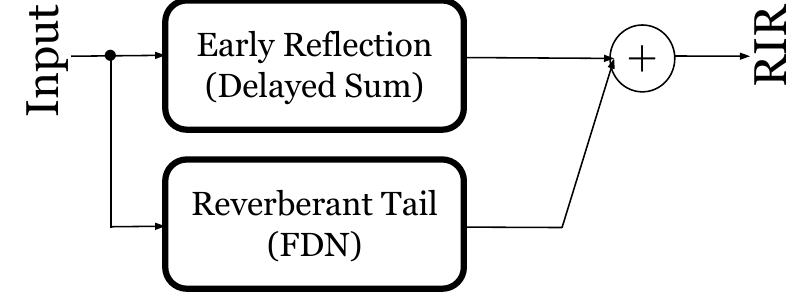}
  \caption{The employed design for applying the early reflections (top, delayed sum network), reverberant tail (middle, feedback delay network), and room impulse response (bottom, overall network) in the $Z$ domain. The $Z$ exponent represents delay. For binaural synthesis, the HRIR specific to the direction of each path  should be applied for the early reflections, and in a general direction towards the source for the reverberant tail.\label{fig:diagram}}
\end{figure}
\vspace*{0pt}
\section{Proposed Method}
\label{sec:method}\vspace*{0pt}
We employ separate processing architectures for the early reflections and the reverberant tail, and combine their outputs.
\vspace*{0pt}
\subsection{Early Reflections}
\label{early}\vspace*{0pt}
Due to the  inherently sparse nature of the early reflections, we employ a delayed sum network as depicted in Fig.~\ref{fig:diagram}—top. The parameters $b_i$ and $K_i$ represent path gains and delays, respectively. These parameters do not require learning since their values are directly derived from the early reflections in the given RIR, which can be obtained through acoustic measurements, computational simulation, or design specifications. This approach integrates conveniently with the HRIR approximation architecture in \cite{gerami2025efficient}.
\subsection{Reverberant Tail}\vspace*{0pt}
If we were to adopt the same approach as for early reflections, we would end up with thousands of coefficients\footnote{A span of 0.5 s would result in $\sim 24\times 10^3$ coefficients.}, and the cost of convolution in either time or frequency domain would be very large.  Since the RIR tail has a decaying pattern, and as humans are only sensitive to characteristics such as $C$ or $T_{30}$, we propose a FDN with a feedback gain of $\alpha$ as depicted in Figure~\ref{fig:diagram}—middle. Our FDN consists of a sum of 16 feedback loops\footnote{This is more stable than a single feedback loop with 16 sums.}, which represent a delayed sum of decaying exponentials. The goal is to tune the FDN coefficients, $\alpha$, $\beta_i$, and $\kappa_i$, which represent the decay rate ($\alpha < 1$), scale, and delay of each exponential, so that the overall network combined with the early reflections depicted in Figure~\ref{fig:diagram}—bottom matches the metrics ($C$, $D$, $CT$, $T_{30}$) of the target RIR. This network results in $149$ floating-point operations (FLOPs) per input; $85$ for the early reflections and their HRIRs, and $63$ for the reverberant tail. In comparison, convolution requires thousands of FLOPs, and the Fourier transform requires hundreds of FLOPs and introduces latency.
\vspace*{0pt}
\subsection{Proposed Differentiable Optimization}\vspace*{0pt}
Given target metrics $C,\, D,\, CT,\, T_{30}$ and early reflections, the network parameters $b_i,\, K_i$ are directly mapped as explained in Section~\ref{early}. To find the FDN parameters, we need to solve the constraints below to find $\alpha,\,\beta_i,$ and $\kappa_i$. We denote the early reflection and the reverberant tail FDN outputs as $I(t)$ and $J(t)$.
\begin{align}
    &\dfrac{\int_0^{50\text{ms}}I^2(t)\,dt + \int_0^{50\text{ms}}J^2(t)\,dt}{\int_0^{\infty}I^2(t)\,dt + \int_0^{\infty}J^2(t)\,dt} = 10^C
    \label{eq1}\\
    &\dfrac{\int_0^{80\text{ms}}I^2(t)\,dt + \int_0^{80\text{ms}}J^2(t)\,dt}{\int_0^{\infty}I^2(t)\,dt + \int_0^{\infty}J^2(t)\,dt} = D\\
    &\dfrac{\int_0^{\infty}tI^2(t)\,dt + \int_0^{\infty}tJ^2(t)\,dt}{\int_0^{\infty}I^2(t)\,dt + \int_0^{\infty}J^2(t)\,dt} = CT\\    &J(T_{30}) = 10^{-3}I(0)\\ \label{eq4}
    &J^2(t) = \sum_{i=1}^{16}\beta_i\,\alpha^{t-\kappa_i}\,u(t-\kappa_i),\quad u(t) \coloneqq \text{Step}.
\end{align}
Since $I(t)$ is known, the $\int I^2(t)\,dt$ are constants that do not affect optimization. For readability, we will omit them. The constraints as currently formulated are difficult to solve. Instead, we develop a convex optimization approach to approximate the solution. To begin we find $\int J^2(t)\,dt$
\begin{align}
    \int_{0}^{T} J^2(t)\,dt &= \int_{0}^{T}(\sum_{i=1}^{16}\beta_i\,\alpha^{t-\kappa_i}\,u(t-\kappa_i))^2\, dt\\
    &= \sum_{i=1}^{16}(\int_{\kappa_i}^{\kappa_{i+1}}(\sum_{j=1}^{i}\beta_j\alpha^{-\kappa_j})^2\alpha^{2t}\, dt\\
    &= \gamma\, \sum_{i=1}^{16}((\alpha^{2k_{i+1}}-\alpha^{2k_{i}})(\sum_{j=1}^{i}\beta_j\alpha^{-\kappa_j})^2)\\
    &= \gamma\, \sum_{i=1}^{16}(\alpha^{2k_{i+1}}-\alpha^{2k_{i}})\lambda_i.\\
    \lambda_i = &(\sum_{j=1}^{i}\beta_j\alpha^{-\kappa_j})^2 \quad \kappa_{17} = T, \quad \gamma = \dfrac{1}{2ln\,\alpha}
\end{align}
Similarly, we find $\int tJ^2(t)\,dt$
\begin{align}
     \int_{0}^{T} tJ^2(t)\,dt 
     = \gamma\sum_{i=1}^{16}((\kappa_{i+1}-\dfrac{1}{2})\alpha^{2k_{i+1}} - (\kappa_{i}-\dfrac{1}{2})\alpha^{2k_{i}})\lambda_i.
\end{align}
We now rewrite the constraints as
\begin{align}
    &\dfrac{\sum_{i=1}^{15}(\alpha^{2k_{i+1}}-\alpha^{2k_{i}})\lambda_i + (\alpha^{100\text{ms}}-\alpha^{2k_{16}})\lambda_{16}}{\sum_{i=1}^{15}(\alpha^{2k_{i+1}}-\alpha^{2k_{i}})\lambda_i + (0-\alpha^{2k_{16}})\lambda_{16}} = 10^C\\
    &\dfrac{\sum_{i=1}^{15}(\alpha^{2k_{i+1}}-\alpha^{2k_{i}})\lambda_i + (\alpha^{160\text{ms}}-\alpha^{2k_{16}})\lambda_{16}}{\sum_{i=1}^{15}(\alpha^{2k_{i+1}}-\alpha^{2k_{i}})\lambda_i + (0-\alpha^{2k_{16}})\lambda_{16}} = D\\
    &\dfrac{\sum_{i=1}^{16}((\kappa_{i+1}-\dfrac{1}{2})\alpha^{2k_{i+1}} - (\kappa_{i}-\dfrac{1}{2})\alpha^{2k_{i}})\lambda_i}{\sum_{i=1}^{15}(\alpha^{2k_{i+1}}-\alpha^{2k_{i}})\lambda_i + (0-\alpha^{2k_{16}})\lambda_{16}} = CT\\
    &\sum_{i=1}^{16}\beta_i\,\alpha^{T_{30}-\kappa_i} = 10^{-6}I(0),
\end{align}
where we assumed  $\underset{i\leq16}{\forall} \kappa_i < 50\text{ms}$ and $\kappa_{17} = \infty$. Multiplying both sides by the denominator, and then subtracting the right side from the left we arrive at
\begin{align}
    &(1-10^C)(\sum_{i=1}^{15}(\alpha^{2k_{i+1}}-\alpha^{2k_{i}})\lambda_i - \alpha^{2k_{16}}\lambda_{16}) \nonumber\\
    &+ \alpha^{100\text{ms}}\lambda_{16}) = 0 \coloneqq \ell_1\\
    &(1-D)(\sum_{i=1}^{15}(\alpha^{2k_{i+1}}-\alpha^{2k_{i}})\lambda_i - \alpha^{2k_{16}}\lambda_{16}) \nonumber\\
    &+ \alpha^{160\text{ms}}\lambda_{16}) = 0 \coloneqq \ell_2\\
    &(1-CT)(\sum_{i=1}^{15}((\kappa_{i+1}-\dfrac{1}{2})\alpha^{2k_{i+1}} - (\kappa_{i}-\dfrac{1}{2})\alpha^{2k_{i}})\lambda_i \nonumber\\
    & - (\kappa_{16}-\dfrac{1}{2})\alpha^{2k_{16}}) = 0 \coloneqq \ell_3\\
    & \sum_{i=1}^{16}\beta_i\,\alpha^{T_{30}-\kappa_i} - 10^{-6}I(0) = 0 \coloneqq \ell_4.
\end{align}
In practice, we are dealing with discrete time. As a result, attempting to solve for $\kappa_i$ would lead to integer programming with a vast solution space. Instead, we set the $\kappa_i$ on a logarithmic space between 0-50\text{ms}. We should emphasize that this will not affect the existence of a viable solution.
\par The $\ell_1,\, \ell_2,\, \ell_3,\, \ell_4$ loss functions are convex with respect to $\alpha$ and $\beta_i$. As a result, a solution can be simply found through gradient descent using a differentiable programming implementation such as Pytorch\cite{Paszke2017-qu} or JAX \cite{Hollander2020-fl}.
\begin{align}
    \alpha,\, \beta_i &= \underset{\alpha, \beta_i}{\text{argmin}}\,\,\,\ell_1 + \ell_2 + \ell_3 + \ell_4,\quad 0 < \alpha, \beta_i < 1.
\end{align}
\section{Experiments}
\label{sec:exp}
To evaluate our approach, we synthesize a real-world RIR~\cite{GTU-RIR} using our algorithm. The synthesis process involves several steps: ideally, we would separate the early reflections from the reverberant tail. However, this separation is challenging in practice due to reverberations and potential overlap between higher-order reflections and the first two orders. Therefore, we identify the peaks with the highest magnitudes within the designated time frame as the early reflections, and define the remaining signal as the reverberant tail. Our implementation consists of two networks: a delayed sum network and an FDN as depicted in Figure~\ref{fig:diagram}. For the delayed sum, we directly map the early reflection to the network's parameters. For the FDN, we find the coefficients so that the desired psychoacoustic metrics (Equations~\ref{eq1}-\ref{eq4}) match the actual RIR metrics. We should emphasize that providing the actual RIR is not necessary; rather, only the early reflections and the desired metrics are required.

\par
\begin{figure}[h!] 
  \centering
  \includegraphics[width=\columnwidth]{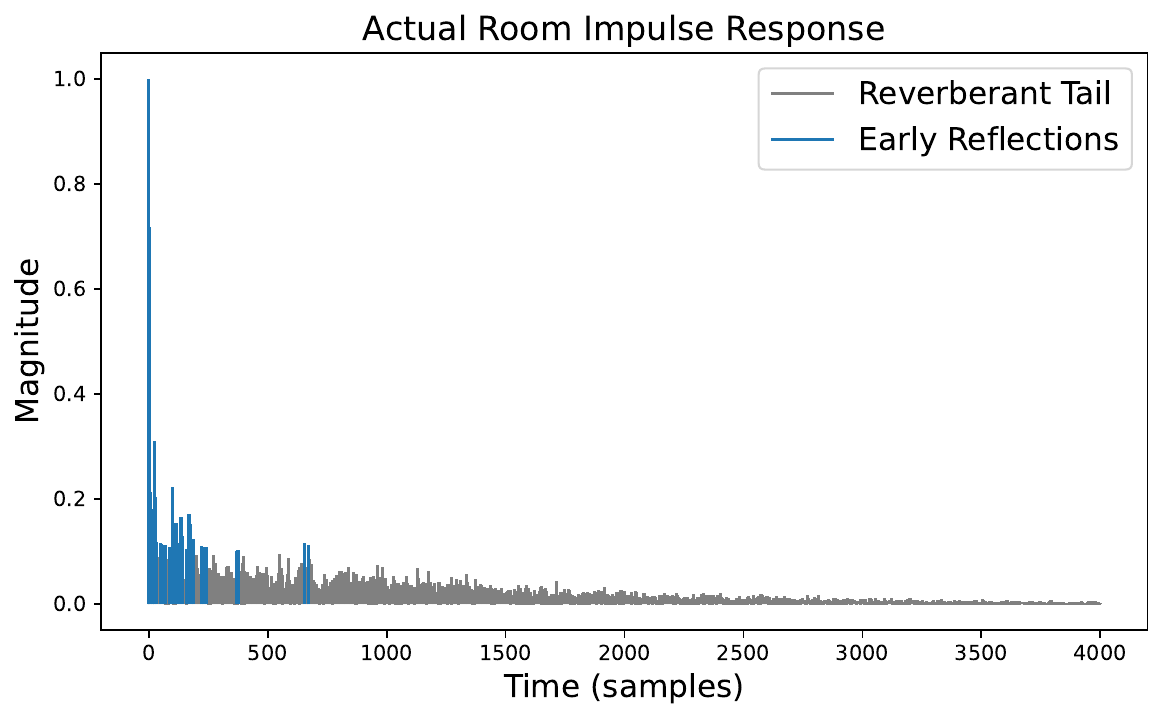} 
  \includegraphics[width=\columnwidth]{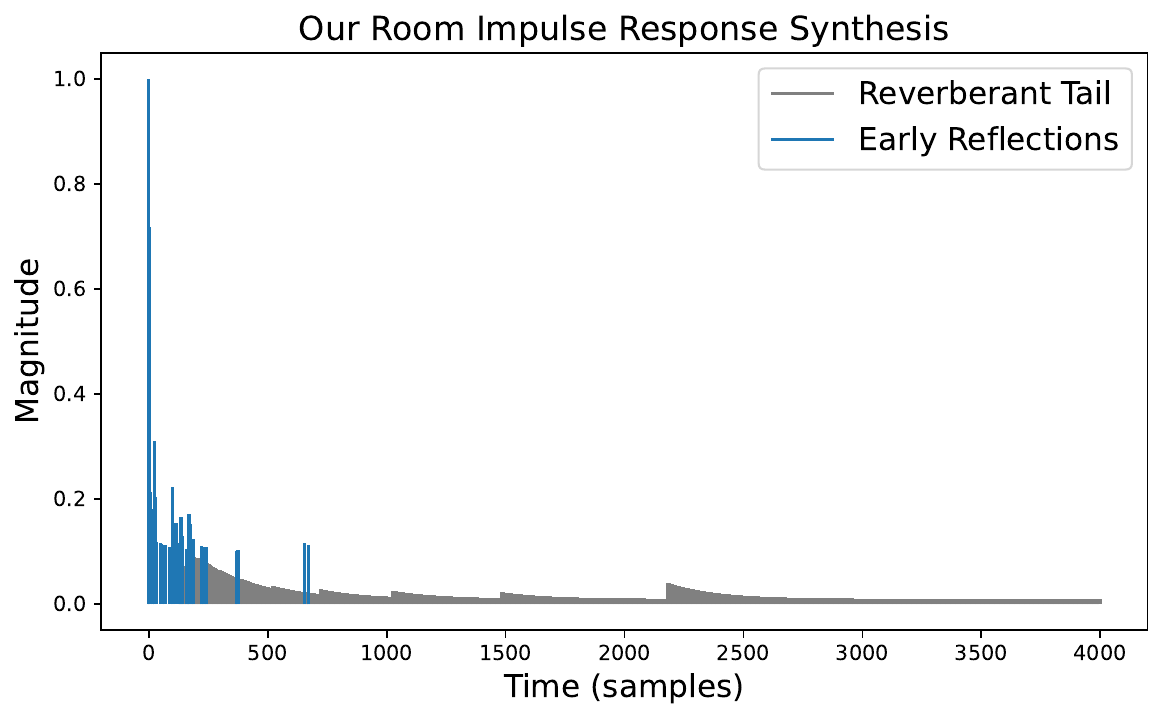} 
  \includegraphics[width=\columnwidth, trim=0mm 0mm 0mm 0mm, clip=true]{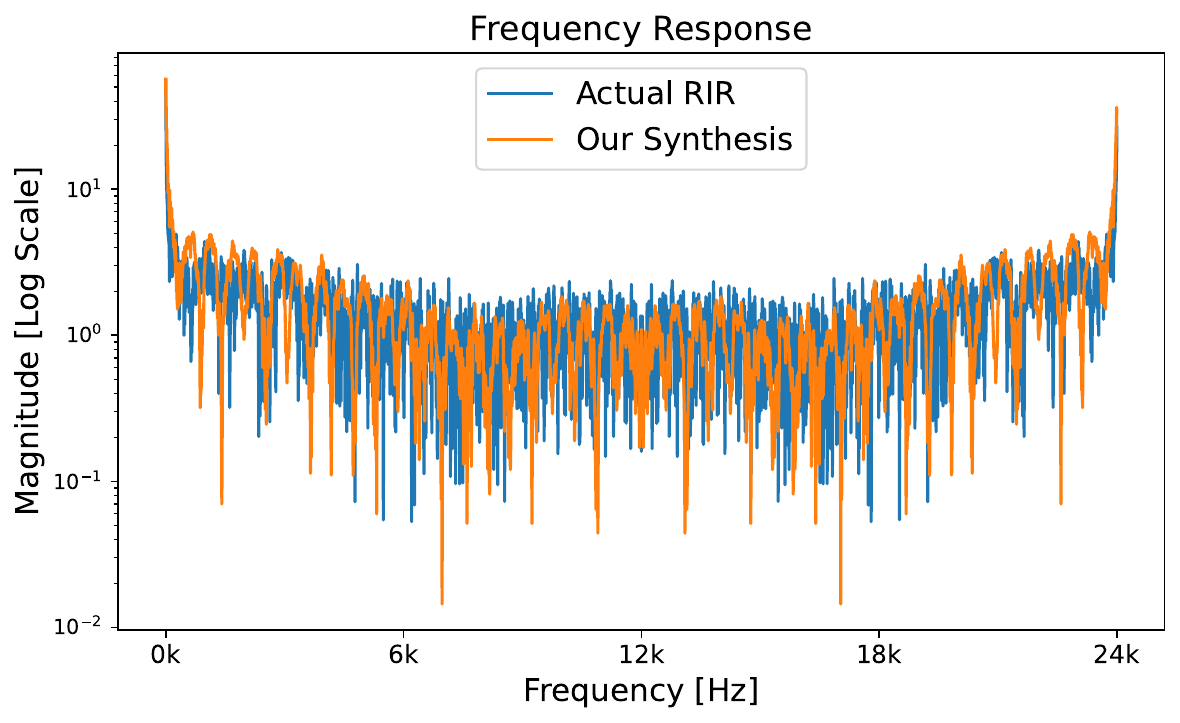} 
  \caption{Magnitude of the actual room impulse response (top) and our synthesized room impulse response (middle) for the first $4000$ time steps. They both share the early reflections, and their reverberant tails follow an exponential decay. Their discrete Fourier transforms (bottom) have the same characteristics as well.}
  \label{fig:result}
\end{figure}

 Figure~\ref{fig:result}-top shows the actual RIR for a small classroom with a sampling rate of $48\,$KHz, and Figure~\ref{fig:result}-middle our synthesized RIR. The designated early reflections from the actual RIR is the same as the early reflections from our synthesis, and both reverberant tails follow an exponential decay pattern. While the specific values of the reverberant tails differ, the learned parameters, obtained through our proposed optimization, result in our synthesis precisely matching the Clarity, Definition, and Center Time metrics of the actual RIR, as demonstrated in Table~\ref{table}. The slight discrepancy in $T_{30}$ can be attributed to our loss function, which is based on the impulse response value at $T_{30}$ rather than the precise time step. As a result, given the small values within that time frame, the loss is small as well, leading to slow convergence. Considering that the difference is $\sim 1$~ms, the effect should be negligible.

\par Looking at the Frequency Response in Figure~\ref{fig:result}-bottom, we can see that the frequency characteristics of the actual RIR are well-captured by our synthesis. This is due to the fact that the identical high-magnitude early reflection components and the shared exponential decay pattern of the reverberant tails. Moreover, our implementation produces a natural, non-metallic sound due to its lack of frequency selectivity. This smoothness is a result of the FDN's structure, which is a sum of decaying exponentials. The Fourier transform of a decaying exponential $f(t) = e^{-kt}$ is $F(\omega) = \frac{1}{k + j\omega}$, a smooth function.

\par The main motivation behind our design is improved computational efficiency. As detailed in Section~\ref{sec:method}, our implementation requires $149$ FLOPs to apply the RIR to a single time step of the input signal. As for convolution based approaches, assuming a window size of $T_{30}$, the computational cost will be $9\times 10^3$ FLOPs\footnote{Convolution with window of $N$ requires $N$ multiplication and additions.} per time step for the RIR. Moreover, $T_{30}$ and the computational cost will increase for bigger rooms. As for Fourier based methods, efficient cyclical based methods~\cite{zotkin2004rendering} require $O((N/W)\log(W) + W)$ FLOPs per time step, where $N$ is the RIR size and $W$ the FFT window size. This approach will also introduce a delay of $W$  since we would have to wait for $W$ time steps to take the FFT. Assuming $N = T_{30}$ and $W = 512$, the computational cost will be $342$ FLOPs\footnote{$N/W$ cyclical windows; each window requires $W\log(W)$ additions and multiplications for FFT and inverse FFT, and $W$ multiplications to apply the transfer function in frequency domain. This will be $(N/W)(\log(W)\times2\times2 + 1)$ FLOPs per input.}. Our implementation achieves $53\times$ and $2.3\times$ reduced computational cost compared to convolution and FFT based approaches without introducing any delay.

\begin{table}[h]
\centering
\resizebox{1\columnwidth}{!}{
\begin{tabular}{@{}lcccccc}
\toprule
& Clarity& Definition& \makecell{Center\\Time}& $T_{30}$&\makecell{Compute\\Eff.\\(Conv.)}&\makecell{Compute\\Eff.\\(FFT)}\\
\hline
\makecell{Actual\\ RIR}& -0.00388& 0.9918& 263.96& 4,735 & -&-\\
\hline
\makecell{Our RIR\\Synth.}& -0.00488& 0.9918& 264.00& 4,248 & 53$\times$&2.3$\times$\\
\bottomrule
\end{tabular}
}
\caption{Comparison of psychoacoustic metrics for the actual room impulse response and our synthesis. Our synthesis matches the metrics while having higher computational efficiency compared to convolution and FFT based approaches.}
\label{table}
\end{table}

\section{Conclusion}
We introduce a computationally efficient feedback delay network (FDN) for real-time room impulse response (RIR) rendering, addressing the computational and latency challenges inherent in traditional convolution and Fourier transform-based methods. Our synthesis results in an RIR that matches the actual RIR's early reflections and psychoacoustic metrics while achieving $53\times$ and $2.3\times$ reduced computational cost compared to convolution and FFT based approaches, and without introducing any delay. When combined with a previous approach to efficiently apply HRIRs to signals using IIR approximations \cite{gerami2025efficient}, we can achieve extremely efficient BRIR filtering and create object based sound in spatial audio on edge devices.

\newpage
\bibliographystyle{IEEEbib}
\bibliography{main.bib}

\end{document}